\documentclass[review]{elsarticle}
\usepackage{float}
\usepackage{geometry}
\usepackage{epstopdf}
\usepackage{hyperref}
\usepackage{color}
\usepackage{amsmath}

\journal{Chaos Solitons \& Fractals}

\begin{document}

\begin{frontmatter}
\title{A novel parameter for nonequilibrium analysis in reconstructed state spaces}

\author[mymainaddress,myfirstaddress]{Wenpo Yao\corref{mycorrespondingauthor}}
\cortext[mycorrespondingauthor]{Corresponding author}
\ead{yaowp@njupt.edu.cn}
\author[mysecaddress]{Wenli Yao}
\author[mymainaddress]{Jun Wang}

\address[mymainaddress]{School of Geographic and Biologic Information, Smart Health Big Data Analysis and Location Services Engineering Lab of Jiangsu Province, Nanjing University of Posts and Telecommunications, Nanjing 210023, Jiangsu, China}
\address[myfirstaddress]{School of Life Science and Technology, University of Electronic Science and Technology of China, Chengdu 611731, China}
\address[mysecaddress]{Department of hydraulic engineering, School of civil engineering, Tsinghua University, Beijing 100084, China}

\begin{abstract}
Kernel methods are widely used for probability estimation by measuring the distribution of low-passed vector distances in reconstructed state spaces. However, the information conveyed by the vector distances that are greater than the threshold has received little attention. In this paper, we consider the probabilistic difference of the kernel transformation in reconstructed state spaces, and derive a novel nonequilibrium descriptor by measuring the fluctuations of the vector distance with respect to the tolerance. We verify the effectiveness of the proposed kernel probabilistic difference using three chaotic series (logistic, Henon, and Lorenz) and a first-order autoregressive series according to the surrogate theory, and we use the kernel parameter to analyze real-world heartbeat data. In the heartbeat analysis, the kernel probabilistic difference, particularly that based on the Kullback--Leibler divergence, effectively characterizes the physiological complexity loss related to reduced cardiac dynamics in the elderly and diseased heartbeat data. Overall, the kernel probabilistic difference provides a novel method for the quantification of nonequilibria by characterizing the fluctuation theorem in reconstructed state spaces, and enables reliable detection of cardiac physiological and pathological information from heart rates.
\end{abstract}


\end{frontmatter}

%\tableofcontents

\section{Introduction}
Complex systems exhibit various features, such as nonlinearity, nonstationarity, and nonequilibria. The nonequilibrium state is the absence or lack of an equilibrium or balance, and numerous concepts and parameters have been proposed to understand nonequilibrium conditions. Fluctuation theorems provide analytical expressions that describe nonequilibrium processes and allow thermodynamic concepts to be extended to finite systems \cite{Sevick2008,Searles2004,Hasegawa2019}. Time irreversibility or temporal asymmetry \cite{Yao2021NS,Costa2005Eq,Cammaro2007}, entropy production \cite{Brunelli2018,Lucia2020,Roldan2010}, and several other mathematical descriptions \cite{Colang2014,Ropke2013} have been formulated to characterize nonequilibrium systems. To extract nonequilibrium features, probability estimation is a necessary part of these quantitative methods, but this is nontrivial for real-world time series, particularly those from complex systems. 

To address the problem of computing the probability distribution, various methods have been developed from different perspectives \cite{Hlava2007}, e.g., those based on maximum likelihood, distance computation, and coarse-grained symbolic transformation \cite{Cammaro2007,Daw2003,Watt2019,Zanin2021}. Among these approaches, model-based linear estimators are not suitable for complex signals with unknown characteristics. Currently, model-free estimators, particularly those based on kernel functions (e.g., the Heaviside function) in reconstructed state spaces are particularly popular \cite{Rojo2018}. Kernel estimation for the state space distance has been widely adopted in informational statistics, such as for entropy estimation \cite{Cang2020,Brechtl2018,Xiong2017,Richman2000}, causality detection \cite{Hlava2007,Schreiber2000T,Marina2007}, and networked connections \cite{Zou2019,Marina2008}. Overall, kernel-based estimators have been used to determine the dynamical complexity of time series in subjects ranging from physics to economics and physiology.

To estimate the probability distribution, the elements in the original series are reconstructed into state spaces, and the probability of the series is calculated indirectly using the distribution of the target spaces. For the Heaviside kernel, the probability distribution of the original process becomes that of the maximum norm differences (i.e., the Chebyshev distance) between space vectors that are within a given tolerance \cite{Hlava2007,Cang2020,Brechtl2018,Xiong2017,Richman2000,Schreiber2000T}. Vectors with spatial distances no larger than the threshold are assumed to have the same structural information and are employed in the probability estimation, while those with differences greater than the threshold are discarded. However, the vectors with distances above the threshold should not be neglected, as there is a possibility that their probability distributions might convey valuable information about the system.

In this paper, we study the information underlying the probabilistic differences between vector distances that fall within and outside the given tolerance. We first introduce the methodology of probability estimation based on the step kernel of space vectors, and propose the kernel probabilistic difference (KPD) parameter. We elucidate the KPD for the nonequilibrium characteristic detection from the perspective of measuring the fluctuation in the Heaviside kernel space. After validating the effectiveness of KPD using the surrogate theory, we conduct a comparative analysis of KPD and the kernel Shannon entropy (KEn) using real-world heartbeat data. The consistency of KPD and permutation time irreversibility in heart rate analysis and the relationship between the kernel method and symbolic ordinal patterns in probability estimation are discussed in detail. Our work provides novel insights into the use of kernel probability estimation and on the quantification of nonequilibrium in reconstructed state spaces.

\section{Method}
Let us introduce the probability estimation based on the Heaviside kernel, and propose the KPD parameter for nonequilibrium detection in reconstructed state spaces.

\subsection{Heaviside kernel probability estimation}
Given the time series $Y(i)=\{x(1),x(2),\ldots,x(i),\ldots,x(L)\}$, multidimensional state spaces are reconstructed according to Eq.~(\ref{eq1}), where $m$ is the dimension and $\tau$ is the time delay.

\begin{eqnarray}
	\label{eq1}
	X_{\tau}^{m}(i)=\{ x(i),x(i+\tau),\ldots,x(i+(m-1)\tau)\}
\end{eqnarray}

We then compute the Minkowski distance between the state vectors $X_{\tau}^{m}(n)$ and $X_{\tau}^{m}(j)$, denoted as $d_{n,j}$ in Eq.~(\ref{eq_2}):

\begin{eqnarray}
	\label{eq_2}
		d_{n,j} =\sqrt[p]{\sum_{s=1}^{m} [x(n+s)-x(j+s)]^{p}}
\end{eqnarray}

According to $p$, the Minkowski distance has three common forms: 1) $p$=1, $d_{n,j}$ is the Manhattan distance, i.e., the summation of absolute differences; 2) $p$=2, $d_{n,j}$ is the Euclidean distance; 3) $p$=$\infty$, $d_{n,j}$ is the Chebyshev distance, the maximum difference among all the distances. Among these measures, the Chebyshev distance, i.e., the maximum norm, is the most common vector distance in probability estimations \cite{Cang2020,Brechtl2018,Xiong2017,Richman2000,Schreiber2000T} for determining whether all differences of values in two state vectors fall inside a given threshold. 

According to Richman and Moorman \cite{Richman2000}, if self-matches are included in measuring the vector distance, the conditional probability in entropy estimation is greater than the unbiased probability. To avoid the bias caused by including self-matches, we employ the sample entropy method \cite{Richman2000}, which does not count self-matches, i.e., $n\neq j$, when estimating probabilities for the analysis of finite datasets using the Chebyshev distance, as given by Eq.~(\ref{eq2}). 

\begin{eqnarray}
	\label{eq2}
		d_{n,j} =\|X_{\tau}^{m}(n),X_{\tau}^{m}(j)\|=\mathop{max}\limits_{1\leq s \leq m}|x(n+s)-x(j+s)|
\end{eqnarray}

The threshold $r$ given by Eq.~(\ref{eq3}), also called the tolerance \cite{Xiong2017,Richman2000,Schreiber2000T}, is set by the control parameter $k$ and the standard deviation (STD), and provides a weighting for the distance of each space vector, where $\bar{x}$ is the mean of the series.

\begin{eqnarray}
	\label{eq3}
	r=k*\sqrt[2]{\sum (x(i)-\bar{x})^{2}/L}
\end{eqnarray}

The Heaviside kernel, i.e., the step kernel, is then applied to determine the relationship between the vector distances and the threshold $r$, as shown by Eq.~(\ref{eq4}).

\begin{eqnarray}
	\label{eq4}
	K=\Theta (d_{n,j}-r)=
	\left\{
	\begin{array}{lr}
		0, \ d_{n,j}>r \\
		1, \ d_{n,j}\leq r
	\end{array}
	\right.
\end{eqnarray}

\begin{figure}[htb]
	\centering
	\includegraphics[width=13cm,height=7cm]{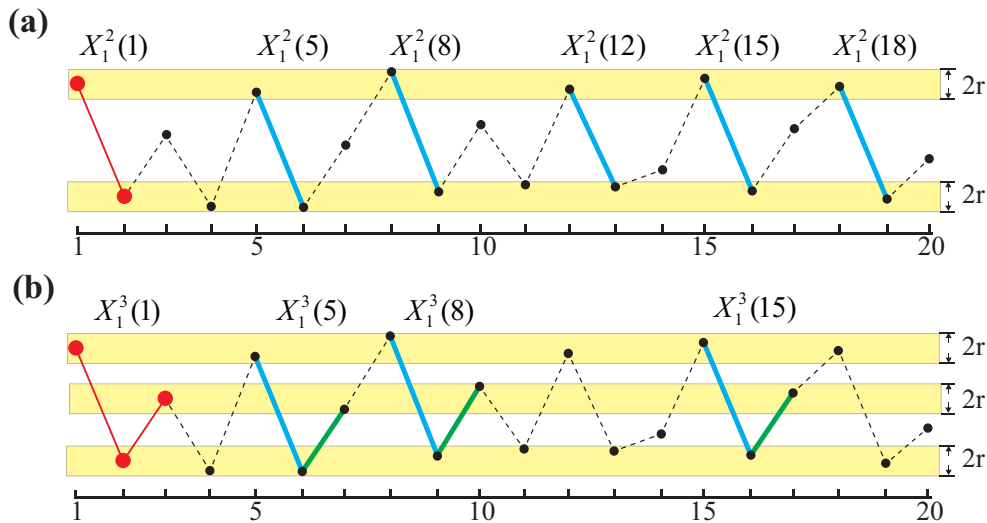}
	\caption{Schematic illustration of kernel transformation for the reconstructed vectors and tolerance. a) When the dimension $m$ = 2, the vectors in bold blue, namely $X_{1}^{2}(5)$, $X_{1}^{2}(8)$, $X_{1}^{2}(12)$, $X_{1}^{2}(15)$, and $X_{1}^{2}(18)$, fall within the tolerance of the first vector, $X_{1}^{2}(1)$. b) When $m$ increases to 3, the third element is connected by the red bold line to the first vector $X_{1}^{3}(1)$, and three vectors, i.e., $X_{1}^{3}(5)$, $X_{1}^{3}(8)$, and $X_{1}^{3}(15)$, have differences with $X_{1}^{3}(1)$ that are no greater than the tolerance. In the vector labels, superscripts denote the dimension $m$, subscripts denote the time delay $\tau$, and values in brackets represent the number of space vectors, e.g., $X_{1}^{2}(15)$ is the 15th vector in phase space with dimension 2 and delay 1.}
	\label{fig1}
\end{figure}

An exemplary illustration of the step kernel transformation for space vector distances with dimensions of $m$=2 and 3 and the threshold $r$ is shown in Fig.~\ref{fig1}.

Figure~\ref{fig1} illustrates the concept of measuring probability distributions that underlies the Heaviside kernel probability estimator. If the largest difference between the corresponding elements in two space vectors is no greater than the threshold $r$, i.e., all differences between two vectors' elements are within the tolerance, the two vectors are assumed to have the same structural information. By calculating the probability distributions of these (assumed same) vectors, we can estimate the approximate probability distribution of the series. This methodology is the kernel-based probability estimation used in most informational parameters \cite{Hlava2007,Brechtl2018,Xiong2017,Richman2000,Schreiber2000T,Marina2007,Zou2019}.

\subsection{KPD in reconstructed state spaces}
Different from most informational parameters, which focus on the vector distances falling within the tolerance, the proposed KPD measures the probabilistic differences of vector distances between given tolerances. Let us introduce the KPD parameter.

The counts of $K$=1 and $K$=0 in Eq.~(\ref{eq4}), i.e., the vector distances that fall within and outside the tolerance, are denoted as $K_{1}$ and $K_{0}$, and their probability distributions are calculated by Eq.~(\ref{eq5}), where the divisor $'L-(m-1)\tau-1'$ is the number of vector distances $d_{n,j}$ of paired vectors with $X_{\tau}^{m}(n)$.

\begin{eqnarray}
	\label{eq5}
	\left\{
	\begin{array}{lr}
		p_{n}^{0}=K_{0}/(L-(m-1)\tau-1)  \\
		p_{n}^{1}=K_{1}/(L-(m-1)\tau-1)
	\end{array}
	\right.
\end{eqnarray}

The probability of $X_{\tau}^{m}(j)$ being within $r$ of $X_{\tau}^{m}(n)$, i.e., $p_{n}^{1}$, is extracted by the step kernel estimation. Different from most information-theoretic approaches that target $p_{n}^{1}$ \cite{Hlava2007,Brechtl2018,Xiong2017,Richman2000,Schreiber2000T,Marina2007,Zou2019}, the KPD index involves both $p_{n}^{1}$ and $p_{n}^{0}$, and particularly their difference. We measure the probabilities of $p_{n}^{1}$ and $p_{n}^{0}$ using Eq.~(\ref{eq6}).

\begin{eqnarray}
	\label{eq6}
	\left\{
	\begin{array}{lr}
		P_{0}=\sum _{n} p_{n}^{0}/(L-(m-1)\tau) \\
		P_{1}=\sum _{n} p_{n}^{1}/(L-(m-1)\tau)
	\end{array}
	\right.
\end{eqnarray}

To quantify probabilistic differences, the popular division-based Kullback--Leibler divergence (also named the relative entropy) and the subtraction-based \emph{Ys} metric \cite{Yao2021NS,Yao2020APL,Yao2019TI} are employed in this paper. The KPD values of $P_{1}$ and $P_{0}$ measured by \emph{Ys} and the \emph{KL} distance are given by Eqs.~(\ref{eq7}) and (\ref{eq8}), respectively, where $P_{1}\geq P_{0}$ (otherwise, exchange the positions of $P_{1}$ and $P_{0}$).

\begin{eqnarray}
	\label{eq7}
	KPD_{Y}= Ys\langle P_{1},P_{0} \rangle = \frac{P_{1}}{P_{1}+P_{0}} \cdot (P_{1}-P_{0})
\end{eqnarray}

\begin{eqnarray}
	\label{eq8}
	KPD_{K}= KL\langle P_{1},P_{0} \rangle = P_{1} log (P_{1}/P_{0})
\end{eqnarray}

In practical computation, the threshold $r$ is usually set to be a fraction of the data STD given by Eq.~(\ref{eq3}), allowing measurements on datasets with different amplitudes to be compared \cite{Xiong2017,Richman2000,Schreiber2000T}. If the threshold value is too large, the probability estimation will be coarse. In the extreme case of very large thresholds, the vector distances all fall within the threshold, making the probability estimation invalid. Small thresholds provide more detailed probability estimations, but the estimates can fail if $r$ is too small. According to empirical analysis, $r$ should generally be 0.2--0.3 times the STD of the dataset so that neither $P_{1}$ nor $P_{0}$ will be zero in the case where KPD$_{K}$ is infinite. 

From a conceptual perspective, KPD quantifies the nonequilibrium feature by measuring the fluctuation theorem from the kernel space. The fluctuation theorem describes some universal statistical properties of nonequilibrium fluctuations and provides analytical expressions for descriptions of nonequilibrium states \cite{Sevick2008}. According to the Evans--Searles fluctuation theorem \cite{Sevick2008,Searles2004,Evans2002}, the dissipation function $\Omega$ of observed trajectories is a dimensionless dissipated energy that is defined by the arbitrary values $A$ and $-A$; their probabilistic difference $p(A)/p(-A)$ describes the asymmetry in the distribution of $\Omega$ over a particular ensemble of trajectories as:

\begin{eqnarray}
	\label{eq_9}
	\frac{p(\Omega_{t}=A)}{p(\Omega_{t}=-A)}=exp(A)
\end{eqnarray}

\begin{figure}[htb]
	\centering
	\includegraphics[width=9.5cm,height=6.5cm]{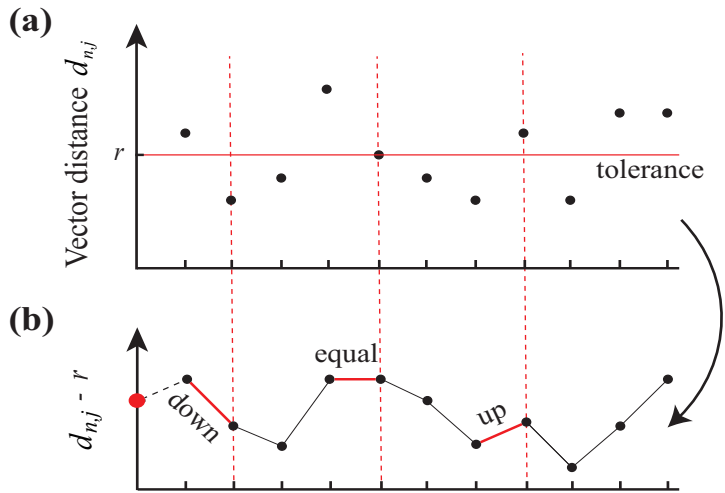}
	\caption{Fluctuations in the differences between vector distances and the tolerance. Vector distances corresponding to the tolerance in a) are transformed into $\Omega= d_{n,j}-r$ in b). $d_{n,j} > r$, $d_{n,j} < r$, and $d_{n,j} = r$ in a) denote up, down, and equality, as illustrated by the red lines in b).}
	\label{fig2_}
\end{figure}

According to the Evans--Searles fluctuation theorem, if we define $\Omega=d_{n,j}-r$ as the space vector distances corresponding to the tolerance, then $d_{n,j} < r$ and $d_{n,j} > r$ represent $-A$ and $A$, and they differ from the original series fluctuations. Thus, KPD$_{Y}$ and KPD$_{K}$ are consistent with the fluctuation theorem as formulated in Eq.~(\ref{eq9}), and they measure the differences between `ups' and `downs' in the transformed series $\Omega=d_{n,j}-r$, i.e., the kernel asymmetry of a given time series, as shown in Fig.~\ref{fig2_}. The dissipation function $\Omega=d_{n,j}-r$ accumulates along the space vector distances corresponding to the tolerance.

\begin{eqnarray}
	\label{eq9}
	KPD_{Y} \propto KPD_{K} \propto \frac{p(d_{n,j}-r=A)}{p(d_{n,j}-r=-A)}=exp(A)
\end{eqnarray}

Therefore, the KPD, which measures the fluctuations of vector distances falling within and outside the tolerance, is a novel parameter for the quantitative nonequilibrium in reconstructed state spaces.

\section{Test results}
In this section, we first generate model series with known characteristics to verify the effectiveness of KPD according to the surrogate theory. We then apply the KPD parameter to analyze real-world heart rate signals.

\subsection{Model series analysis}
Let us first check the effectiveness of KPD according to the surrogate theory. The surrogate method \cite{Lancaster2018,Schreiber2000} formulates a suitable null hypothesis for a process, then generates surrogate data that are consistent with this hypothesis, before finally comparing a discriminating statistic for the original and surrogate datasets to determine whether to reject or accept the null hypothesis. Generally, if the discriminating index of the original process is less than the 2.5th percentile or greater than the 97.5th percentile of the surrogate dataset, the null hypothesis is false and should be rejected; otherwise, the null hypothesis is true and should be accepted.

To test the KPD$_{Y}$ and KPD$_{K}$, we first generate chaotic series using the logistic equation, mathematically written as $x_{t+1}=r \cdot x_{t} (1- x_{t})$, the two-dimensional Henon map $x_{t+1}=1-\alpha \cdot x^{2}_{t}+y_{t}$, $y_{t+1}=\beta \cdot x_{t}$, and the Lorenz system generated from three coupled differential equations, $dx/dt=\sigma (y-x)$, $dy/dt=x(r-z)-y$, and $dz/dt=xy-bz$. A linear series is generated using the first-order autoregressive (AR1) process given by $x_{t+1}=\delta x_{t} + \xi_{t}$, where $\xi_{t}$ is the Gaussian random variable with zero mean and unit variance. For each model series, 500 sets of surrogate data with the same autocorrelations and power spectra are generated using the model-free improved amplitude-adjusted Fourier transform \cite{Schreiber2000,Schrei1996}.

\begin{figure}[htb]
	\centering
	\includegraphics[width=16cm,height=10cm]{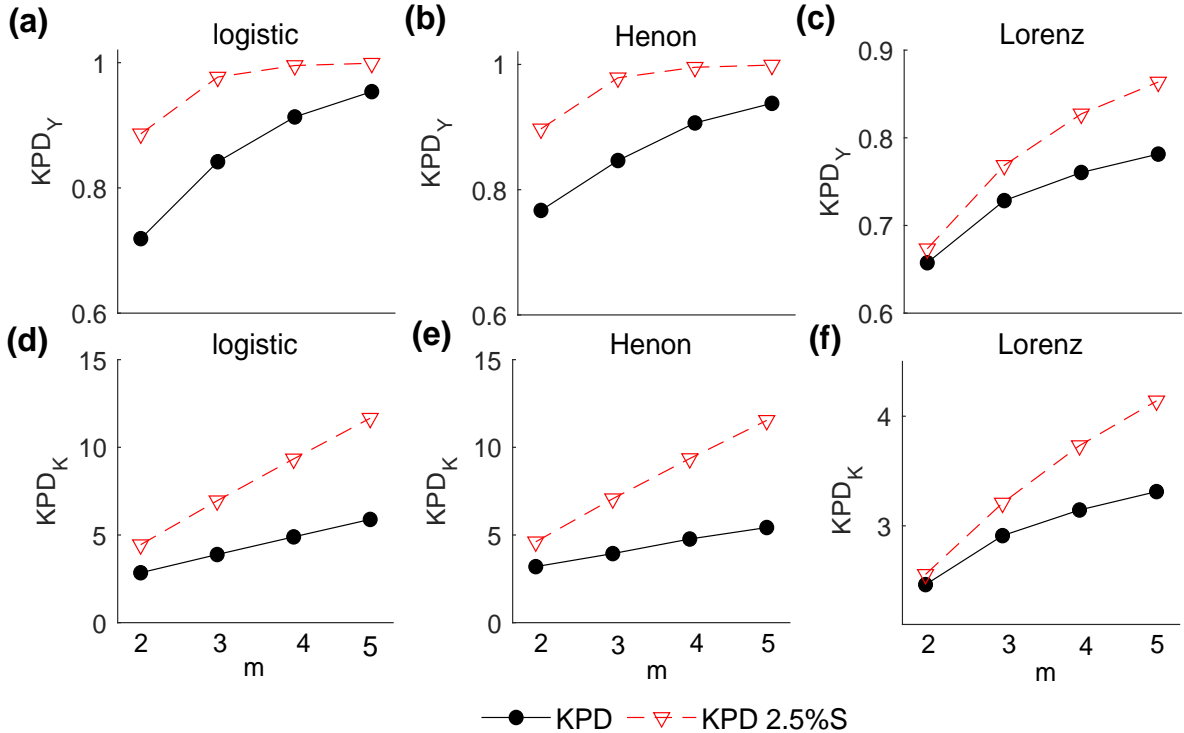}
	\caption{KPD of the logistic, Henon, and Lorenz series and their surrogates. The initial values of the logistic ($r$=4) and Henon ($\alpha$=1.4, $\beta$=0.3) series are $x_{1}$=$y_{1}$=0.01, and those of the Lorenz ($\delta$=10, $r$=28, and $b$=8/3) series are $x_{1}$=0, $y_{1}$=0, and $z_{1}$=1$\times10^{-10}$. The x-components of the Henon and Lorenz systems are employed. The tolerance is given by a $k$ = 0.3. `2.5\%S' denotes the 2.5th percentile of the surrogate data}
	\label{fig2}
\end{figure}

The KPD$_{Y}$ and KPD$_{K}$ values of the three sets of chaotic data and their surrogates with a data length of 6000 are illustrated in Fig.~\ref{fig2}. Those of the AR1 series ($\delta$=0.3) are listed in Table~\ref{tab1} because their \emph{KPD} values are very similar and cannot be effectively differentiated in a plot. In the table, `2.5\%S' and `97.5\%S' denote the 2.5th and 97.5th percentiles of the surrogate data, `Ys3' represents KPD$_{Y}$ with $m$=3, `KL3' represents KPD$_{K}$ with $m$=3, and so on.

\begin{table}[htb]
	\centering
	\caption{KPD of the AR1 model series and its surrogate data.}
	\label{tab1}
	\begin{tabular}{ccccc cccc}
		\hline
		AR1 &Ys2	&Ys3	 &Ys4	 &Ys5	 &KL2	 &KL3	 &KL4 &KL5	 \\
		\hline
		2.5\%S  &0.310	&0.585	&0.762	&0.866	&0.959	&2.077	&3.158	&4.171	 \\
		$KPD$   &0.311	&0.586	&0.764	&0.869	&0.962	&2.082	&3.173	&4.204	 \\
		97.5\%S &0.315	&0.594	&0.770	&0.874	&0.976	&2.123	&3.218	&4.271	 \\
		\hline
	\end{tabular}
\end{table}

According to Fig.~\ref{fig1} and Table~\ref{tab1}, KPD based on both \emph{KL} and \emph{Ys} can reliably detect these model series. In Fig.~\ref{fig1}, the KPD values of the logistic, Henon, and Lorenz series are all smaller than the 2.5th percentile of their surrogates, while in Table~\ref{tab1}, the KPD values of the AR1 data are all between the 2.5th and 97.5th percentiles of the surrogate data. The null hypotheses that the logistic, Henon, and Lorenz series are linear are all false and should be rejected, while the null hypothesis that the AR1 series is linear is true and should be accepted. The test results for KPD in these model series are in line with the fact that chaotic logistic, Henon, and Lorenz series are nonlinear and that the AR1 series is linear.

In these numerical simulations, the three chaotic series all have significantly smaller KPD values than their surrogate data. Taking the logistic series as an example, the thresholds of the logistic function and its surrogate data are all 0.106 when $m$=2. The probability of vector distances falling within the threshold is 0.100, whereas that of the surrogate data is around 0.039. Thus, KPD$_{Y}$ for the logistic series is 0.719, significantly smaller than that of its surrogate data, 0.887. The stochastic surrogate datasets have a greater degree of randomness, and so their vector distances have a larger distribution range and a smaller probability of being within the threshold. Therefore, the chaotic series have significantly smaller KPD values than their surrogate data.

Therefore, as verified by the four model series, KPD$_{Y}$ and KPD$_{K}$ are effective parameters according to the surrogate theory.

\begin{figure}[htb]
	\centering
	\includegraphics[width=12cm,height=7cm]{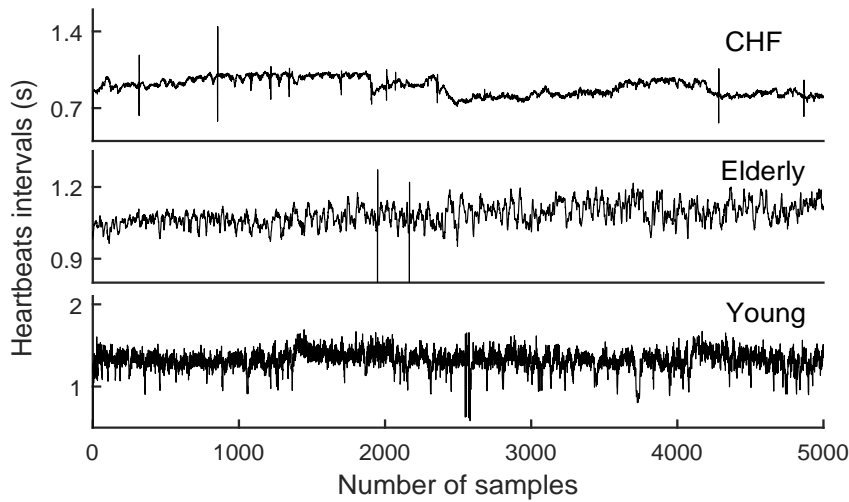}
	\caption{Example time series from the CHF, healthy elderly, and healthy young heartbeats.}
	\label{fig3}
\end{figure}

\subsection{Real-world heartbeats analysis}
Cardiac activity, particularly heartbeats, contains important information about physiological complexity \cite{Yao2021NS,Costa2005Eq,Cammaro2007,Yao2020APL,Perc2005,Billman2011}. In this subsection, we apply KPD to real-world heartbeats obtained from the public PhysioNet database \cite{Goldber2000}.

Congestive heart failure (CHF) is a pathological condition in which the heart has degraded pumping function \cite{Isler2019}, and is a common cardiac disease in older adults. PhysioNet contains two CHF databases, `chfdb' and `chf2db,' which include heartbeats from long-term electrocardiography recordings. We combine these databases and collect the heartbeats of 44 patients with CHF (mean age 55.5$\pm$11.4 years, ranging from 22 to 79). Additionally, the heart rates of 20 healthy young adults (mean age 25.8$\pm$4.3 years, ranging from 21 to 34) and 20 healthy elderly adults (mean age 74.5$\pm$4.4 years, ranging from 68 to 85), both with equal numbers of women and men, were recorded in a resting state with a sinus rhythm while watching the movie `Fantasia.' Similar groups of heartbeats have been widely used to analyze the cardiac dynamics with regard to cardiac regulation \cite{Yao2021NS,Costa2005Eq,Xiong2017,Yao2020APL,Yao2019E}. Exemplary heartbeat intervals of the CHF, healthy elderly, and healthy young subjects are depicted in Fig.~\ref{fig3}.

\begin{figure}[htb]
	\centering
	\includegraphics[width=16cm,height=10cm]{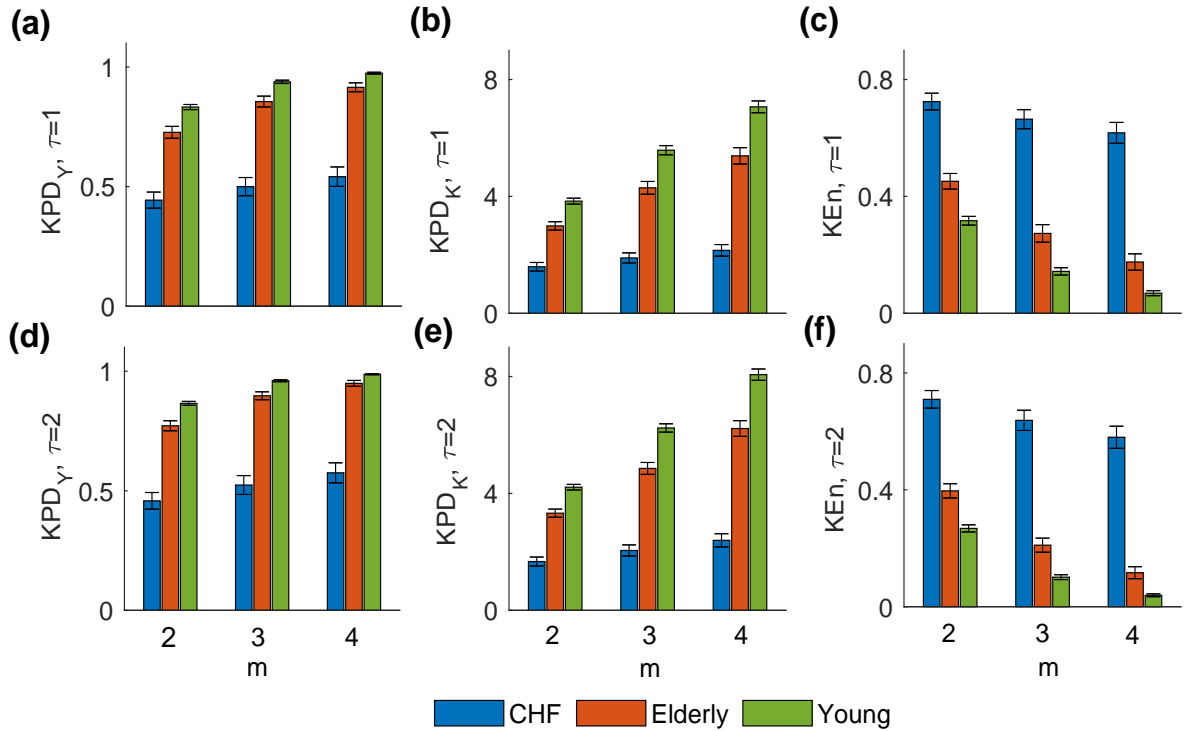}
	\caption{KPD and KEn (mean$\pm$standard error) of the CHF, healthy young, and healthy elderly heartbeats. Threshold computed using $k$ = 0.3 and the dimension varied from 2 to 4; $\tau$=1 and 2. Detailed statistical t-tests for KPD and KEn of heart rates are listed in Table~\ref{tab2}.}
	\label{fig4}
\end{figure}

As a comparison, the Shannon entropy (the classical index measuring informational complexity) for the Heaviside kernel probabilities, denoted as KEn and given by Eq.~(\ref{eq10}), was also employed. The KPD and KEn values of the heartbeat data are shown in Fig.~\ref{fig4}, and the $p$ values of statistical t-tests for the KPD and KEn values of the heartbeat datasets are listed in Table~\ref{tab2}.

\begin{eqnarray}
	\label{eq10}
	KEn= -P_{0}log P_{0}-P_{1}log P_{1}
\end{eqnarray}

\begin{table}[htb]
	\centering
	\caption{Statistical t-tests for KPD and KEn values of the three groups of heartbeats. Independent sample t-tests were employed to determine the differences of each pair of heartbeat datasets, and one-way analysis of variance (ANOVA) was used for the three sets of heartbeats.}
	\label{tab2}
	\begin{tabular}{ccccc ccccc}
		\hline
		&Ys2	&Ys3	&Ys4	&KL2	&KL3	&KL4	&KEn2	&KEn3	&KEn4 \\
		\hline
		C-Y,$\tau$=1	&4.3e-15	&6.9e-15	&1.2e-13	&8.6e-19	&2.3e-22	&1.7e-22	&1.7e-18	&8.7e-21	&1.1e-19 \\
		C-E,$\tau$=1	&5.4e-9	&3.7e-11	&1.7e-11	&8.3e-9	&2.5e-11	&2.6e-13	&4.1e-9	&3.5e-12	&4.8e-14 \\
		E-Y,$\tau$=1	&3.7e-4	&2.1e-3	&5.5e-3	&2.2e-5	&2.6e-5	&2.1e-5	&8.0e-5	&4.8e-4	&1.3e-3 \\
		C-E-Y,$\tau$=1	&3.9e-13	&4.0e-14	&2.2e-13	&1.2e-16	&8.1e-23	&1.0e-25	&7.0e-16	&2.7e-19	&1.3e-19 \\
		\hline
		C-Y,$\tau$=2	&3.7e-15	&3.5e-14	&1.4e-12	&7.6e-21	&1.2e-25	&1.2e-26	&3.1e-19	&8.4e-20	&4.3e-18 \\
		C-E,$\tau$=2	&1.3e-10	&5.6e-12	&2.3e-11	&8.3e-11	&1.3e-13	&1.3e-14	&3.6e-11	&8.9e-15	&1.2e-15 \\
		E-Y,$\tau$=2	&1.6e-4	&1.3e-3	&6.2e-3	&6.0e-6	&2.0e-6	&2.1e-6	&3.3e-5	&2.6e-4	&1.4e-3 \\
		C-E-Y,$\tau$=2	&5.2e-14	&4.7e-14	&1.4e-12	&8.4e-19	&4.8e-25	&8.8e-26	&2.1e-17	&5.6e-20	&3.5e-19 \\
		\hline
	\end{tabular}
\end{table}

According to Fig.~\ref{fig4}, KPD$_{Y}$ and KPD$_{K}$ exhibit consistent results, with the healthy young heartbeats displaying the greatest degree of nonequilibria and the CHF heartbeats having the lowest degree of nonequilibria. However, the KEn analysis for the three heartbeat datasets is totally contrary to KPD, with the CHF data giving the highest KEn and the healthy young data give the lowest KEn. According to Table~\ref{tab2}, the three kernel nonlinear parameters are all significantly discriminative of each pair of heartbeat datasets: the $p$ values of statistical tests of the CHF and healthy heartbeats are all less than 5.4$\times10^{-9}$, those of the healthy young and elderly heartbeats are less than 0.007, and those of the three groups of heart rates are less than 1.4$\times10^{-12}$. Statistically, KPD$_{K}$ provides the best discrimination between each pair of heartbeat datasets, with KEn performing better than KPD$_{Y}$ but slightly worse than KPD$_{K}$. It is also apparent that the parameters in the phase space have no significant effects on KPD and KEn for the three groups of heartbeats, suggesting that kernel-based methods are highly insensitive to the parameters of the phase space construction.

From a physiological perspective, the KPD parameters reflect the cardiac dynamics of complexity loss theory in heart rates. According to the general dynamical model for pathophysiology, i.e., the theory of complexity loss with aging and disease \cite{Goldber2002}, healthy physiological systems reveal a type of complex variability along with various nonlinear interactions under certain conditions, and this type of nonlinear complexity decreases and even breaks down with aging and disease. As for the cardiac dynamics in heartbeats, the healthy young heartbeats should have the greatest degree of complexity, whereas the elderly and particularly the CHF heartbeats should have less-complex dynamics because of the reduced adaptive capabilities of cardiac regulation. The wisdom of complexity loss theory has been widely proved using entropy measures \cite{Xiong2017}, irreversibility \cite{Yao2021NS,Costa2005Eq,Yao2020APL,Yao2019E}, and fractality \cite{Goldber2002,Ivanov1999}, and is now further confirmed by \emph{KPD}.

Interestingly, the contradictory findings regarding KPD and KEn are consistent with our previous report on the Shannon entropy and quantitative time irreversibility \cite{Yao2020APL}. Mathematically, the Shannon entropy characterizes complexity as the mean of information, and so smaller probabilistic differences between $P_{1}$ and $P_{0}$ will produce larger values of KEn (i.e., the complexity, unpredictability, and randomness), but smaller values of KPD$_{Y}$ and KPD$_{K}$. According to complexity loss theory \cite{Goldber2002}, the CHF heartbeats have the least nonlinear dynamics, and the probabilistic difference between the space vectors' differences within and outside the tolerance is small, i.e., the lowest KPD and the largest KEn. In contrast, the healthy young heartbeats have the greatest cardiac dynamics as well as the largest 0-1 difference in their kernel transformation. Therefore, the healthy young heartbeats produce the largest KPD and the smallest KEn. These contradictory findings about KPD and KEn are in line with complexity loss theory with respect to cardiac dynamics. More importantly, they further suggest that KPD and KEn characterize different, and even contradictory, aspects of complex systems \cite{Yao2020APL}.

According to the test results, KPD$_K$ and KPD$_Y$ are both effective parameters for nonequilibria detection. KPD, particularly KPD$_{K}$, significantly discriminates between the three groups of heartbeats and is consistent with complexity loss theory. Overall, the method of measuring the probabilistic difference between the vector distances falling within and outside the tolerance for nonequilibria detection is effective and reliable.

\section{Discussion}
The KPD parameter has been proven to be effective and reliable for the detection of nonequilibria in reconstructed state spaces. Other issues requiring further discussion include the characteristics of KPD in quantitative nonequilibria and the relationship between the kernel transformation and permutation in probability estimation.

According to the KPD and KEn results with real-world heartbeats, from both mathematical and physiological perspectives, KPD has strong similarities with time irreversibility \cite{Yao2020APL}, suggesting their connection is particularly interesting and should be discussed in detail. Based on the statistical definitions of time irreversibility, we need to calculate the probabilistic difference between symmetric vectors or between forward--backward vectors; however, kernel functions are not suitable for quantifying time irreversibility, because the vectors in the original time series have been transformed and the corresponding vectors cannot be matched after the kernel transformation. In our previous report on quantitative time irreversibility and amplitude irreversibility \cite{Yao2021NS}, we made the assumption that we could directly measure the nonequilibrium characteristic of complex systems, rather than only focusing on the reversible or symmetric features. We also postulated that it would be possible to quantify nonequilibrium characteristics according to the probabilistic differences in transformed signals, such as kernel functions, network constructions, and so on. KPD is a solution for quantitative nonequilibria from the perspective of kernel transformation. The consistent results of KPD and time irreversibility regarding the heartbeat data and the link between KPD and the quantitative fluctuation theorem confirm our assumptions and show that KPD provides an alternative for the quantification of nonequilibrium.

Another issue concerns the roles of the kernel method and permutation in informational probability estimation. The kernel transformation is based on the distances between space vectors, while permutation is a kind of coarse-grained symbolic transformation \cite{Watt2019,Zanin2021,Bandt2002}. In the original proposition of the transfer entropy, Schreiber \cite{Schreiber2000T} adopted the step kernel of maximum distance to estimate the transition probabilities. Later, considering the great demands on data, fine-tuning parameters, and sensitivity to noise of the traditional methods, Staniek and Lehnertz \cite{Staniek2008} proposed the permutation transfer entropy, which is robust and computationally fast. In this paper, we have found that the KPD and KEn values obtained from heartbeat data are not significantly affected by the parameter selection in the state space construction. However, our previous studies have shown that the permutation probabilistic differences and permutation irreversibility are affected by these parameters \cite{Yao2019E,Yao2019PDP}, suggesting the advantage of kernel transformation, i.e., its high reliability. Moreover, KEn and KPD have advantages over permutation-based parameters in other aspects. For instance, equal values might produce misleading and even completely incorrect results, which could be significant in heartbeat analysis where there are large numbers of equal values \cite{Yao2020APL,Yao2019E,Yao2019PDP,Bian2012}. Additionally, equal values in time irreversibility have significant physical implications, as self-symmetric vectors and their permutations are time reversible. The existence of forbidden permutations (order patterns that do not exist) \cite{Yao2021NS,Yao2020APL,Yao2019E} means that the division-based Kullback--Leibler divergence is not appropriate for measuring the probabilistic differences for the permutation time irreversibility. This issue does not affect the KPD parameter, because it is possible to ensure that $P_{0}$ or $P_{1}$ is nonzero by adjusting the threshold $r$. At the same time, we should note that the probability estimation based on the kernel transformation of the vector distance involves two rounds of calculation of vector differences; therefore, it consumes a larger amount of time than the coarse-graining of ordinal patterns. In short, the kernel method and permutation have their own characteristics and particular applications, and should be selected accordingly to guarantee more reliable and accurate results.

\section{Conclusions}
To conclude, we proposed a novel effective KPD parameter for nonequilibrium analysis in reconstructed state spaces by measuring the 0-1 difference of the Heaviside kernel transformation. The KPD conceptually measures the fluctuation theorem from the kernel state space instead of the original process for the quantitative nonequilibrium, and its effectiveness is verified by model series according to the surrogate theory. In heartbeats analysis, the KPD confirms the theory of complexity loss in heartbeats from aging and diseased subjects, and its contradictory results to the KEn improve our understanding about different aspects of the complex cardiac system. Moreover, the consistent findings of the KPD with time irreversibility highlights its physical significance in broadening the quantitative nonequilibrium.

\nocite{*}

\bibliography{mybibfile}% Produces the bibliography via BibTeX.

\end{document}